# Populating the Digital Space for Cultural Heritage with Heritage Digital Twins

Franco Niccolucci [1,*], Achille Felicetti [2] and Sorin Hermon [3]

1   PIN, Prato, Italy; franco.niccolucci@gmail.com
2   PIN, Prato, Italy; achille.felicetti@gmail.com
3   STARC, The Cyprus Institute, Nicosia, Cyprus; sorin.hermon@cyi.ac.cy
*   Correspondence: franco.niccolucci@gmail.com

**Abstract:** The present paper concerns the design of the semantic infrastructure of the digital space for cultural heritage as envisaged by the European Commission in its recent documents. Due to the complexity of the cultural heritage data and of their intrinsic inter-relationships, it is necessary to introduce a novel ontology, yet compliant with existing standards and interoperable with previous platforms used in this context, such as Europeana. The digital space organization must be tailored to the methods and the theory of cultural heritage, briefly summarized in the introduction. The new ontology is based on the Digital Twin concept, i.e. the digital counterpart of cultural heritage assets incorporating all the digital information pertaining to them. This creates a Knowledge Base on the cultural heritage digital space. The paper outlines the main features of the proposed Heritage Digital Twin ontology and provides some examples of application. Future work will include completing the ontology in all its details and testing it in other real cases and with the various sectors of the cultural heritage community.

## 1. Introduction

With its recent *Recommendation of 10/11/21 on a common European data space for cultural heritage* [1] the EU Commission has set the foundations and defined the features of a framework supporting and hosting the digital transformation of cultural heritage, i.e. the process and the results of leveraging digital technologies to transform how heritage organisations operate and deliver value. This section will introduce the general concept of cultural heritage to explain what the objective of such digital transformation should be, and which is the perspective and the methodology of the heritage community. Other relevant EU policy documents include [2], [3] and [4].

The concept of cultural heritage and of its safeguard, conservation, and valorisation, has evolved in time since the 18th century. We may date back to that period the creation of the present most important museums (Uffizi, 1737; Hermitage, 1764; British Museum, 1753; Louvre, 1793); the first large-scale archaeological excavations (first systematic excavation of Pompeii, 1749); and, in general, the interest in antiquities by antiquarians (English Society of Antiquarians chartered in 1780). The "Family Pact" by the last member of the Medici family and Electress of the Palatinate Anna Maria Luisa de' Medici, i.e. her testament signed in 1737 and eventually leading to the creation of the Uffizi Gallery, stated that all heritage assets in Florence should remain there as "ornament of the state, for the utility of the public and to attract foreigners' curiosity". This is perhaps the first definition of the mission of cultural heritage institutions.

Since then, an innumerable set of statements, discussion, papers, and books has been dedicated to this topic. It has been the subject of international charters and has informed heritage practice and methodology. Hence, it would be impossible to address here such theoretical background in detail. However, it has been clearly outlined in the EU report *Innovation in Cultural Heritage Research* [4], which describes the so-called cultural *regimes* and the progress of the notion of cultural heritage. The following list is derived from such report.

- The *First Regime* (circa 1860 – 1960s) is determined by national and local heritage conservation regulations and it lasts until the codification of international cultural heritage protection. It is mainly based on national initiatives, with international charters starting to appear in its final stage. Actions address the physical conservation and enhancement of monuments, sometimes allowing substantial modification as in Viollet-Le-Duc 'reconstructions', later deprecated in the Venice Charter (1964) for philological fidelity to the original; and the exhibition of the atemporal beauty of art masterpieces. The public is mainly formed by 'visitors', guided by the experts, archaeologists, museum curators and other heritage professionals.
- The *Second Regime* (1960s – 1990s) corresponds to the institutionalisation of cultural heritage as an international value. UNESCO and its related institutions (ICOM, ICOMOS) are the chief actors. Diverse communities with their cultures can express and see recognized their own testimonies, including intangible ones. Deeper understanding of heritage by visitors is fostered with various means. The role of heritage in understanding past and present societies is enhanced.
- The *Third Regime* (late 1990s to present) corresponds to the renewed institutionalisation of cultural heritage characterised by its expansion in terms of concepts, significance, number and diversity of heritage assets and elements. A significant feature is the involvement of communities and their contribution to the definition and construction of cultural heritage, sometimes described as co-creation.

The development of digital technologies in the present century is producing deep effects on our lifestyle, on our work and on our communication attitude. Nevertheless, in most cultural heritage-related activities the digital component has not been interiorized by institutions and professionals, using it as an appliance to perform the same tasks and to achieve the same goals as before. It is now high time for fully incorporating the digital component into heritage methods and practice, moving the concept of heritage towards Heritage 4.0, i.e. the *Fourth Regime*, characterized by the digital transformation effects on cultural heritage activities, institutions and people, both professionals and citizens.

Such digital transformation is not simply the deployment of advanced digital technologies to aid traditional practices and business models. Careful consideration needs to be given to how the traditional core of the organizations can benefit of digitization and explore and capture new ways of creating value. This new approach will progressively add to and partially replace the traditional one, without superseding the values and goals of heritage institutions, summarized in the current ICOM definition of a museum [6], (Art. 3 Section 1) as "a non-profit, permanent institution in the service of society and its development, open to the public, which acquires, conserves, researches, communicates and exhibits the tangible and intangible heritage of humanity and its environment for the purposes of education, study and enjoyment". This definition dates back to 1974 and was updated several times, the last one in 2017. At the 2019 Tokyo ICOM Extraordinary General Assembly a new one has been presented, but it raised a still unresolved controversy and is currently suspended. Our approach seems to fit also with the discussed new definition, which stresses the social values of cultural heritage.

Since 2017, when this definition was updated, social networks have become commonplace, and the digital component is an integral part of our daily life. Digital identities – legal and functional/social ones – are replacing real ones under many regards. Communication is social, quick, often superficial, and broadcast to many. This impacts both on the traditional definition of museums, here used as representatives of the whole heritage sector, and on the one newly proposed including multivocality and social aspects in the museum mission. People increasingly speak digital. Talking and listening to them needs to establish novel communication channels, in which Creative Industries play a substantial role together with 'transformed' heritage professionals and institutions. Moreover, the nowadays available advanced digital technologies for cultural heritage offer incredible opportunities for fulfilling other aspects of heritage institutions mission such as conservation. They need to become part of the heritage methodology, on a par with more traditional ones.

To be up to this new role, digital technologies must support access to accumulated knowledge and be capable of improving their contribution to cultural heritage. Technologies are of no value by themselves, even if advanced: it is when they are used to help making processes related to heritage knowledge, conservation, and communication more effective, and to deliver a better experience, that real value is created.

Such digital evolution requires tools to operate on digital artefacts, which form and populate the digital space of cultural heritage, conceived as a meta-space where digital artefacts duplicate real ones; and through services, commanded by humans or activated by some kind of artificial intelligence (AI), which reproduces heritage processes such as curation, conservation and preservation.

The present paper intends to contribute to the definition of such digital artefacts, to outline how they may be (digitally) manipulated and propose some exemplary applications among the many that the move to digital will enable. To define such artefacts, it is necessary to be aware that heritage comes in very different shapes. They range from immovable property, such as historic buildings, sites, landscapes, environments; to movable ones, such as works of art or technology and science. They also include immaterial resources such as music, performances, traditions, practices, representations, expressions, knowledge, skills – as well as the instruments, objects, and cultural spaces associated with them and recognized as such by a community or group of people. Such aspects are intertwined with each other, linking for example a building to its content and its history. To denote each heritage item, we will use the term *heritage asset*, which may refer to any kind of heritage, either tangible, both movable and immovable, or intangible.

Digital models of heritage assets must therefore be able to encompass all such diversity and incorporate the relationships among the various components of the assets they represent. They must also be able to react to external impulses, reproducing the effects these would have on their real counterpart. This digital framework is already in use in several manufacturing sectors [7]. News in the press report that every Tesla car has a digital duplicate – with which it continuously dialogues – checking the proper functionality of the vehicle and its parts. Bioinformatics allows to describe, study and make *in silico* experiments on complex objects such as proteins and viruses. The design and engineering of spacecrafts are digitally tested in simulations on digital models: actually, this was the first domain in which this approach was extensively introduced by NASA in 2010, being often impossible to reproduce the conditions of space travels. The digital model used in all such cases is called a *digital twin*, a virtual representation that serves as the real-time digital counterpart of a physical object. The digital twin concept has made its way also in the official documents of the European Commission, for example in the Commission's Communication *Shaping Europe's digital future* [5], which envisages a digital twin of the Earth. Such digital twin is also quoted as a substantial component of the Data Space for Smart Communities initiative within the Digital Europe programme.

The same approach is being applied in a domain closer to cultural heritage, the building industry, incorporating in the design of buildings not only the geometry of the object being modelled, but also the information about its physical and functional characteristics, e.g. the material; services such as electricity, plumbing and so on; and the construction process, management and control. This methodology is called Building information Modelling (BIM). In BIM applications such information is attached on a 3D model. The digital representation of architectural, structural and all other elements is based on an international metadata schema, an ISO standard: ISO (2018) 16739-1:2018 [8], called Industry Foundation Classes (IFC). The IFC schema is a standardised data model that codifies the identity (name, machine-readable unique identifier, object type or function), the characteristics or attributes (such as material, colour, and physical properties) and relationships (including locations and connections) of objects (as columns or slabs), abstract concepts (performance, costing), processes (installation, operations), and people (owners, designers, contractors, suppliers, etc.). With this approach, a 3D CAD model becomes the pivot for searching and the preferential entry point for accessing the knowledge about an asset. Among others, the Centre for Digital Built Britain (CDBB), a partnership between the Department for Business, Energy & Industrial Strategy (BEIS) of the UK Government and the University of Cambridge, has started the application of digital twins for the built environment in UK on behalf of the UK government, publishing in a report [9] a summary of the principles a digital twin must comply with, which they call the *Gemini Principles*.

In the literature there has been a very active thread proposing to use BIM as the entry point to access heritage documentation and as the main constituent of the digital twin, renaming this technology as Heritage BIM (HBIM). Google scholar lists about 1400 papers on the subject in 2020 and 2021 only, ranging from the illustration of individual case studies to more general ones. Many different proposals have been advanced to enrich the foundation classes of an HBIM model to incorporate the necessary information concerning shape and materials. All such proposals share a common approach derived from the architectural/civil engineering origin of the model: they are all based on 'enriched' 3D models. With this approach, however, the field of application of digital twins is restricted to immovable physical assets and even for them it does not consider all the features that are relevant in a cultural heritage context, for example the intangible value of a monument. The next sections propose an improved approach which addresses such limitations.

## 2. Results

### 2.1 The definition of the Heritage Digital Twin

The new model proposed in this paper builds on the existing digital twin concept to extend it the to the whole domain of cultural heritage, in its diverse facets of immovable, movable and immaterial/intangible aspects. It reformulates what seems to be a partially distorted perspective totally relying on the object shape and architectural characteristics, derived from the building applications developed so far. The current model is usually regarded as a very rich BIM model, making it unsuitable for movable heritage, which has a material component without the features considered in BIM models, and especially for intangible or born-digital heritage, which lack a physical component at all. Instead, in our approach, we will develop a semantic model of the digital twin where the knowledge about each asset is organised in a knowledge graph, altogether being regarded as the digital twin of the heritage asset. This leads to the following definition:

*The **Heritage Digital Twin (HDT)** of a **Heritage Asset (HA)** is the digital representation of the complex of knowledge about such heritage asset, organised according to a specific ontology, called **the HDT ontology**.*

The HDT includes, but is not limited to, 3D models of the asset. Thus, visual models such as point-cloud ones or HBIM models may be important components of a Heritage Digital Twin but have no privileged status compared to other attributes. Their role ranges from primary (but not exclusive) for architectural assets, to modest or non-existing for intangible assets. From the user's perspective, the information contained in the HDT ontology can be mapped to the IFC schema and produce a HBIM view on HDT via software, in real time or generating permanent HBIM models. The other way round is also feasible, by transforming the HBIM data to HDT content via the inverse mapping of the above-mentioned one. Since all the IFC classes and properties will have corresponding HDT classes and properties, both transformations are lossless. Thus, the HDT of a heritage asset may be considered an extension of the HBIM model, which maps one-to-one to a part of it: but it is indeed much more.

On the other hand, the HDT approach is not only a clever way of organizing information for retrieval, allowing to access data for consultation by a human agent from any point of the knowledge graph that documents the heritage asset. Information may also be used for direct machine processing by means of simulation models integrated with the knowledge base. Thus, messages from sensors or from external big data systems may trigger automated actions. For example, as explained in greater detail in section 3, a weather threat alert coming from a (Big Data) meteo forecast system connected with the heritage knowledge base will launch a software service that retrieves and analyses the heritage assets located in the threatened area and identifies the ones at flood risk, for example comparing the height of their location with the expected level of the flood water. Such system may activate automated actions, such as raising barriers, and alert human actors to intervene to mitigate the adverse effects of this natural disaster, according to the different nature of the assets and of their content as resulting from the collection of all the relevant data provided in HDT, for example moving paintings upstairs.

*2.2 Introducing the Heritage Digital Twin ontology*

The following description of classes and properties of the HDT ontology aims at being readable also by non-technical but digitally aware heritage professionals and researchers; it is also intended to give a preliminary insight into the ontology. A more technical description is provided in section 3, including a summary table. The ontology is described through the explanation of (some of) its classes, i.e. the concepts it organizes by grouping similar instances, i.e. individual occurrences; and by illustrating properties, i.e. the relationships between such concepts. To visually identify them, HDT ontology class and property names will be indicated in *italic*, with class names capitalized. The inverse property of a property will be parenthesized in the definition: thus, the property name *is part of (has part)* means that statements such as "X *is part of* Y" and "Y *has part* X" have the same meaning and are the same property, only swapping the subject with the predicate.

Notice that the same words used for classes and properties may be used in the discourse: in such cases they will not be italicized. As customary in semantics, class names are used in the singular only; when used as plain words, names may obviously be in the singular or plural as required by the context.

All HDT classes are considered as subclasses of an overarching one, *Heritage Entity*, which has no instances. The pivot concepts are the class *Heritage Asset*, corresponding to actual heritage assets of any nature (physical, both movable or immovable, immaterial or born digital), and the class *Heritage Digital Twin*, to indicate the whole of the digital information pertaining to *Heritage Asset*. The *Heritage Digital Twin* is related to the *Heritage Asset* via the property *is digital twin of (has digital twin)*. We will start describing the hierarchical structure of *Heritage Asset* and then introduce the one for *Heritage Digital Twin*.

Any instance of the *Heritage Asset* class may be related to other instances of the same class, for example being a part of them, such as the tower of a castle, or forming with them a collection or a more general asset, as the paintings of an art gallery. Another recent example is the new UNESCO World Heritage site "The Great Spa towns of Europe" [10], which is made of eleven towns, each one developed around a spa, far from each other but belonging to a common cultural framework. Each one will have its own *Heritage Asset* instance, and there will also be a collective *Heritage Asset* instance representing the whole UNESCO World Heritage site.

Information about the parts forming an asset is relevant also to the whole, as it describes important details such as, for example, the material of which the part is made. Thus, parts of a heritage asset may be considered as assets as well, and correspondingly the HDT of the whole will incorporate the HDTs of all the parts of the entire heritage asset which are identified as heritage asset on their own. For example, a church, i.e. the 'main' heritage asset, has many parts such as chapels, paintings, architectural components, furniture, and so on, each one a heritage asset on its own and with an own heritage digital twin. The heritage digital twin of the church is then the assemblage of all such heritage digital twins, plus features concerning the whole, for example the style, the cult, the architect and so on. HAs being parts of an HA are related to the whole by the property *is part of (has part)*, which produces a similar property *is digital part of (has digital part)* linking the corresponding HDTs to the HDT of the whole. A more generic connection than part-whole is described instead by the property *is related to (relates)*, relating heritage assets to other assets. Being symmetric, this property coincides with its inverse.

The instances of all classes *have* an *identifier*, i.e. a code attached to them such as an inventory number; and may *have* one (or more) *Title*, often according to language, e.g. "Mona Lisa" (English), "Monna Lisa" or "La Gioconda" (Italian), "La Joconde" (French) etc., all referring to the same famous Leonardo's painting. Thus, the properties *is identified by (identifies)* and *is titled (titles)* have *Heritage Entity* as domain, since identifiers may be defined also for digital twins, e.g. their URL, as well as names if one likes to do so for digital artefacts.

The *Heritage Document* class includes all the documentation items pertaining to a *Heritage Asset*. Such documentation may consist in digital or analogic objects such as printed or handwritten documents, old analogic photos, drawings, and so on; or digital, either born digital like digital photos or digitised from analogic one.

The property linking *Heritage Asset* to *Heritage Document* is the property *is documented in (documents)*, which may apply to the whole asset or to specific parts of it.

Among others, we may distinguish among the following ones:

- *3D model*, resulting from any of the various techniques available such as 3D scanning, wireframe modelling and so on. Practice will assess if it is more convenient to distinguish among them introducing different types, e.g. 3D point-cloud model, 3D CAD model and so on.
- *Imagery*, such as photos and videos, but also special imagery such as X-ray images, spectra of chemical and physical analyses, and so on. Among them, particularly relevant are the Virtual reality (VR) and Augmented Reality (AR) models, other types of visual digital artefacts pertaining to *Heritage Asset*. Both VR and AR models rely on 3D models of the related heritage asset, but may add or remove parts of it, or require further digital input as in AR, so they should be catalogued separately from 3D models. 3D models may correspond to actual objects – artefacts or built structures – or to conceptual ones, often representing the reconstruction of what is presumed to be the original configuration (and often, the use) of the reconstructed object. Such models are named Virtual Reconstructions and are a commonplace in archaeology to communicate to researchers or, more frequently, to the public, the interpretation of past appearance. In general, Virtual Reality (VR) models enable virtual visits, while models incorporating the present appearance of heritage assets. i.e. Augmented Reality (AR) models can be viewed only on site, as they need the real time acquisition of the current asset appearance.

More types/subclasses may be introduced according to needs. For example, it may be worth considering *Conservation Document*, data about the asset conservation both in terms of the documentation of past interventions, the materials, and the analyses carried out on it. It consists in text files, numeric files (e.g. the results of analyses), images, videos, and special 3D objects, for example the results of tomography. In general, different types of models may generate a subclass if they have special properties that apply only to them. Otherwise, it is simpler to consider their *Type* only, associated via the property *has type (is type of)*.

A *Heritage Asset* pertaining to a tangible asset *is located in (is location of)* a *Place*. Such location may have different levels of precision, as a determinate position or area, a town, a region, and so on, for example "Room 1 of the Uffizi Museum", "Athens", "Cyprus". It may also vary in time, if the related asset is moved elsewhere. A concept similar to location may be considered also for some intangible assets, as is done for some members of the UNESCO Intangible Heritage Representative List [11]. For others, it is intrinsic to the intangible asset to be on the move or to have no location, for example music. Thus, the (important) location property domain does not coincide with all the *Heritage Asset* class and is somehow different between tangible and intangible heritage: for the former it defines where the asset is located, while for the latter it indicates where the asset manifests itself.

Therefore, the Heritage Asset class must split into the two subclasses *Tangible Asset* and *Intangible Asset*. For the former, *is located in (is location of)* specifies where the asset is placed. For the latter, the sister property is *happened at location (was location for)*. Both have *Heritage Location* as the range. Defining the location of a tangible asset may have various degrees of difficulty: if it is easy for most of them, in some cases it relies on research as there is no physical evidence confirming the location. This is the case of battlefields, for instance, when no finds or traces exists. An example is the Cannae site, where the battle (216 BC) between Romans and Carthaginians took place during the Second Punic War. Historical reports by Polybius and Titus Livius are available and enable a trustworthy identification of the location.

A *Heritage Asset* has many a *Heritage Story* that are *associated* to it. A *Story* includes any kind of witness related to the asset: it can be a narrative, a historical source, a popular attribution, co-created content and so on. A very special case concerns a literary itinerary such as Leopold Bloom's route through Dublin in James Joyce's *Ulysses*, or – in a light-hearted perspective – real or almost imaginary places featured in popular novels or TV series, for example crime series as George Simenon's *Commissaire Maigret*, in Paris; or Andrea Camilleri's *Inspector Montalbano*, based in southern Sicily; both generating a substantial flow of 'cultural' tourism. In such cases, it is the *Heritage Story* that creates the *Heritage Asset*. Such 'fake' cultural heritage is discussed in detail in [12].

In general, a *Heritage Story* relates tangible heritage assets to their intangible components and to their reference communities. They are therefore of paramount importance also for the asset physical conservation and the safeguard of its intangible value. Before starting a conservation intervention or the evaluation of an activity from the heritage conservation perspective, it is necessary to consider the impact on the intangible component especially when the impact on the tangible one is irrelevant. For example, fast food shops are often banned from historic centres, although their visual impact may be negligible; there are, instead, provisions to preserve the permanence of historic shops. Locating a MacDonald's in the basement of a historic palace would be unthinkable even if the building statics is unaffected. Ignoring these intangible aspects as it happens in BIM and HBIM models is a serious shortcoming for any heritage application of such approaches and one of the main reasons to propose a generalisation as the HDT ontology. A very nice example concerns Orthodox sacred icons. The devotion to a particular icon is manifested by lighting candles in front of it, which in time causes the blackening of the painting. Thus, the icon blackening level is the evidence of the believers' devotion and a confirmation of its religious value. Therefore, cleaning it would damage such intangible value: a common and much valued restoration practice for paintings would have an unexpected adverse effect if this intangible component is not considered. There are of course stories about intangible heritage as well, often in a much greater amount than for the tangible one.

To clarify the difference between a *Heritage Document* and a *Heritage Story*, let us give a more precise explanation of both.

A *Heritage Document* consists of data produced by a dedicated activity, finalised to produce knowledge about one or more heritage assets. Such production activities include research, management, conservation and restoration, and display of the assets. In some cases, such activities have also an interest independent from the heritage they address, for example as regards the technology or the materials used.

Documents span over many types of data. They include, among others: research reports and data; all kinds of imagery, from B&W photos to 3D models, drawings and maps; video and other recordings, for example interviews with artists and architects as well as with people and communities. They also include historical reports and descriptions, both ancient and modern, e.g. from Vasari's "*Vite*" (*The Lives of the Most Excellent Painters, Sculptors, and Architects*, 3rd edition 1568) to a recent paper on *Art History*.

The activity of producing a *Heritage Document* is called *Heritage Documenting*. It is characterised by the intentionality of creating new knowledge. We deliberately avoid using the term "documentation" in such class names as it has an ambiguous meaning, the activity and its result: "Documentation (i.e. the activity) produces documentation (the outcomes)". The term can be disambiguated only by the context, obviously unavailable in an isolated name.

On the other hand, a *Heritage Story* is the account of facts about a *Heritage Asset*, including but not limited to descriptions based on documents and on the interpretation of such documents. They are usually formulated in an attractive and accessible way to facilitate visitors' understanding and stimulate their curiosity. Frequently they avail of communication techniques such as drawings, physical or digital reconstructions, and increasingly use VR and AR technology.

The level of factualness of a story may vary from interpretation based on research and told as a story, to a somehow arbitrary reconstruction, and to a mostly imaginary legend. For example, the historical existence of King Arthur is debated among historians, probably with no conclusive fact on his real existence due to the lack of historical documents and the doubts existing on the value of mediaeval texts. Conclusions on this regard are disputed as it is widely (but not unanimously) considered as legendary by scholars. Locating Arthur's birthplace and possibly his royal court at Tintagel Castle in Cornwall is nowadays refused by historians. Nevertheless, such stories continue to be told about such early mediaeval ruins excavated in the 20th century, probably as they attract tourists more fascinated by Arthur's legend and the link to the Arthurian cycle of literary legends concerning him, his wife Guinevere, his sword Excalibur and the Knights of the Round Table, rather than by factual accounts about a sixth century small castle and trading point, of which very little remains. The activity of creating and disseminating a story is known as storytelling and thus we will name the class including such activities with the same name, *Storytelling*.

In conclusion, stories are an integral component of the heritage framework as they may make it more understandable, interesting, and attractive for the public. They are distinct from documents, as they do not create – nor aim to create – new knowledge on the assets they concern but rather organise it into a coherent account mixing research results with interpretation based on research at a variable degree. Instead, a *Heritage Document* is aimed at creating new knowledge and is made using a scholarly methodology based on factual evidence, deduction and inference, regardless of the correctness of the premises and the conclusions. In the case of intangible heritage, the border between a *Story* and a *Document* is blurred because the story may be part of the asset value. In this case, we would distinguish between them according to the intentionality of the production activity – research for *Documenting* and explanation, valorisation, communication for *Storytelling*. Regarding to the cultural heritage digital space as a source of content for creative industries, stories with their visual apparatus are perhaps the most valuable information concerning heritage. Creative Industries also produce new stories, which might be linked to the heritage assets concerned.

The association between a heritage asset and its heritage digital twin induces properties relating the latter to the various attributes of the former: digital objects, e.g. a 3D model, related to a heritage asset, are also components of the digital twin of that asset. For example, the property *has visual documentation (is visual documentation of)* for *Heritage Asset* establishes a property *includes visual documentation (visual documentation is included in)* for the related *Heritage Digital Twin*; *is documented in (documents)* for *Heritage Asset* establishes *includes document (document is included in)* for its twin; and so on. Therefore, such induced properties are shortcuts, i.e. they are equivalent to the combination of *is digital twin of* and the corresponding heritage asset property: for instance *Heritage Digital Twin includes document* is a shortcut for *Heritage Digital Twin is digital twin of Heritage Asset* (which) *is documented in Heritage Document*. Since the heritage digital twin is the complex of all information, all such twin properties are subproperties of a superproperty *has as component (is component of)* having *Heritage Digital Twin* as domain.

To conclude this discursive introduction to the Heritage Digital Twin ontology, we notice that the level of ontology definition is still initial and needs further work.

Nevertheless, the approach seems promising as it takes into account aspects relevant for cultural heritage data which are instead ignored by other models. It supports the construction of a data space more functional than relying on other simpler knowledge organization schemes that overlook several important features. A preliminary assessment of the HDT ontology, summarized below, shows that it is suitable for heritage assets as resulting from the above summary description of its classes and properties.

*2.3 The semantics of the HDT ontology*

This section will describe the HDT ontology in a more formal way. Classes will be indicated with a label formed by HC (for heritage class) followed by a progressive number accompanying the class name. e.g. *HC1 Heritage Entity*. Likewise, properties will be denoted by HP and a number, preceding the property name. As before, class and property names will be italicized, and class name also capitalized.

We note that the HDT ontology is a compatible extension – more precisely, an application profile – of the CRM conceptual reference model [13] for heritage documentation, an ISO 21127:2014 standard. As such, it makes the cultural heritage data space interoperable with thousands of repositories of heritage documentation which adopt CRM-compliant data models, from which information can be directly retrieved. Moreover, with this approach the HDT ontology is developed to be compatible with EDM, the Europeana Data Model [14] – actually to be an extension of it to accommodate with a broader use – thus enabling the direct transfer of information to and from the current Europeana system. Several classes and properties of the HDT ontology are indeed equivalent to CRM ones. Referred CRM classes and properties will be denoted by the prefix *crm:*, which applies also to other CRM extensions such as CRMdig, the CRM extension for digital objects; CRMarchaeo, the extension to archaeological investigations; CRMba, the one for building archaeology; CRMsci, the one for heritage science, and so on. Such extensions are characterized by different encoding for the labels, which for instance include E and P for the base CRM, D and L for CRMdig, and so on: thus, class and property labels directly identify the relevant extension they belong to. Reference to such extension is available on the already mentioned CIDOC CRM site [13]. We also notice that the definition of the HDT ontology builds on pre-existing work that has already developed, or is currently developing, parts re-usable for it, mainly in the ARIADNE [15] and in the 4CH [16] EU projects.

For space reasons, the following lists include only the HDT essential classes and properties, with synthetic descriptions. They are intended to give an idea of the HDT ontology just illustrate the data space structure.

2.3.1 Class definitions

*HC1 Heritage Entity*
Subclass of: *crm:E1 CRM Entity*
Description: This class comprises all things in the universe of discourse of cultural heritage data. It has no instances. Its use is mainly to define general properties that are inherited by all its subclasses, i.e. all the classes of the HDT ontology, for example *crm:P1 is identified by* or *crm:P2 has type*.

*HC2 Heritage Asset*
Subclass of: *HC1 Heritage Entity*, *crm:E77 Persistent Item*
Description: Actual heritage asset of any nature: physical, both movable and immovable, immaterial, or born digital.

*HC3 Heritage Digital Twin*
Subclass of: *HC1 Heritage Entity*, *crm:D1 Digital Object*
Description: Collection of all the digital objects pertaining to an *HC2 Heritage Asset*.

*HC4 Heritage Document*
Subclass of: *HC1 Heritage Entity*, *crm:E31 Document*
Description: A piece of the collection of all documents concerning an *HC2 Heritage Asset*.

*HC5 Heritage Documenting*
Subclass of: *HC1 Heritage Entity*, *crm:E7 Activity*
Description: The activity of collecting or digitizing documents concerning an *HC2 Heritage Asset*.

*HC6 Imagery*
Subclass of: *HC5 Heritage Document*, *crm:D1 Digital Object*
Description: a *HC5 Heritage Document* consisting in digital imagery of any kind about an *HC2 Heritage Asset*.

*HC7 3D Model*
Subclass of: *H5 Heritage Document*, *crm:D1 Digital Object*
Description: an *HC5 Heritage Document* consisting of a 3D (digital) model of the heritage asset. AR and VR models are related by an appropriate function to the place where they are to be used.

*HC8 Heritage Event*
Subclass of: *HC1 Heritage Entity*, *crm:E5 Event*
Description: Any external event related or susceptible to have an impact on an *HC2 Heritage Asset*.

*HC9 Heritage Condition State*
Subclass of: *HC1 Heritage Entity*, *crm:E3 Condition State*
Description: Condition of an *HC2 Heritage Asset* caused by an *HC8 Heritage Event* as documented in an *HC4 Heritage Document*: for example, the condition of "destroyed" following an earthquake event.

*HC10 Heritage Event Model*
Subclass of: *HC1 Heritage Entity*, *crm:D1 Digital Object*
Description: Digital model of any *HC4 Heritage Event* susceptible to have an effect on heritage assets. It may range from plain description to simulation software, and is not part of the *HC3 Heritage Digital Twin*. It is used, for example, in simulations.

*HC11 Heritage Identifier*
Subclass of: *HC1 Heritage Entity*, *crm:E42 Identifier*
Description: Identifier for an *HC1 Heritage Entity*, e.g. the DOI for digital items.

*HC12 Heritage Title*
Subclass of: *HC1 Heritage Entity*, *crm:E35 Title*
Description: A title under which an *HC2 Heritage Asset* is known.

*HC13 Heritage Location*
Subclass of: *HC1 Heritage Entity*, *crm:E53 Place*
Description: The location an *HC2 Heritage Asset*.

*HC14 Heritage Story*
Subclass of: *HC1 Heritage Entity*, *crm:E89 Propositional Object*
Description: Account about an *HC2 Heritage Asset*.

*HC15 Storytelling*
Subclass of: *HC1 Heritage Entity*, *crm:E7 Activity*
Description: Activity of creating an *HC14 Heritage Story*.

*HC16 Tangible Asset*
Subclass of: *HC2 Heritage Asset*, *crm:E18 Physical Thing*
Description: Tangible part of an *HC2 Heritage Asset*.

*HC17 Intangible Asset*
Subclass of: *HC2 Heritage Asset*, *crm:E28 Conceptual Object*
Description: Intangible part of an *HC2 Heritage Asset*.

*HP18 Heritage Analogical Document*
Subclass of: *HC4 Heritage Document*
Description: A piece of the collection of all analogical documents of an *HC2 Heritage Asset*.

*HP19 Heritage Digital Document*
Subclass of: *HC4 Heritage Document, crm:D9 Data Object*
Description: A piece of the collection of all digital documents of an *HC2 Heritage Asset*, which form its digital twin.

*HP20 Hosting Service*
Subclass of: *HC1 Heritage Entity, crm:PE5 Digital Hosting Service*
Description: The digital platform or infrastructure on which an *H3 Heritage Digital Twin* is hosted and operated.

*HP21 Hosting Service Provider*
Subclass of: *HC1 Heritage Entity, crm:E39 Actor*
Description: The owner or administrator of the digital infrastructure providing the *HP20 Hosting Service* for operating an *HC3 Heritage Digital Twin*.

2.3.2 Property definitions

*HP1 has digital twin* (*is digital twin of*)
Domain: *HC2 Heritage Asset*
Range: *HC3 Heritage Digital Twin*

*HP2 is documented in* (*documents*)
Domain: *HC2 Heritage Asset*
Range: *HC4 Heritage Document*

*HP3 is related to* (*relates*)
Domain: *HC2 Heritage Asset*
Range: *HC2 Heritage Asset*

*HP4 is part of* (*is formed by part*)
Domain: *HC2 Heritage Asset*
Range: *HC2 Heritage Asset*

*HP5 includes document* (*digital document is included in*)
Domain: *HC3 Heritage Digital Twin*
Range: *HC4 Heritage Document*

*HP6 has story* (*is story about*)
Domain: *HC2 Heritage Asset*
Range: *HC14 Heritage Story*

*HP7 uses story* (*story is used by*)
Domain: *HC3 Heritage Digital Twin*
Range: *HC14 Heritage Story*

*HP8 created document* (*is document created by*)
Domain: *HC5 Heritage Documenting*
Range: *HC4 Heritage Document*

*HP9 creates story* (*is story created by*)
Domain: *HC15 Storytelling*
Range: *HC14 Heritage Story*

*HP10 identifies* (*is identified by*)
Domain: *HC1 Heritage Entity*
Range: *HC11 Heritage Identifier*

*HP11 has component* (*is component of*)

Domain: *HC3 Heritage Digital Twin*
Range: *HC4 Heritage Document*

*HP12 is located in* (*is location of*)
Domain: *HC16 Tangible Asset*
Range: *HC13 Heritage Location*

*HP13 has intangible component* (*is intangible component of*)
Domain: *HC16 Tangible Asset*
Range: *HC17 Intangible Asset*

*HP14 has manifestation event* (*event is manifestation of*)
Domain: *HC17 Intangible Asset*
Range: *HC8 Heritage Event*

*HP15 is manifested by* (*is manifestation of*)
Domain: *HC17 Intangible Asset*
Range: *HC16 Tangible Asset*

*HP16 used document* (*document used for*)
Domain: *HC15 Storytelling*
Range: *HC4 Heritage Document*

*HP17 has visual documentation* (*is visual documentation of*)
Domain: *HC2 Heritage Asset*
Range: *HC6 Imagery*

*HP18 has 3D model* (*is 3D model of*)
Domain: *HC16 Tangible Asset*
Range: *HC7 3D Model*

*HP19 narrates* (*is narrated by*)
Domain: *HC14 Heritage Story*
Range: *HC8 Heritage Event*

*HP20 depicts* (*is depicted by*)
Domain: *HC6 Imagery*
Range: *HC8 Heritage Event*

*HP21 has type* (*is type of*)
Domain: *HC1 Heritage Entity*
Range: *crm:E55 Type*
*[controlled vocabulary, e.g. Getty AAT]*

*HP22 has condition state* (*is condition state of*)
Domain: *HC16 Tangible Asset*
Range: *HC9 Heritage Condition State*

*HP23 tells about* (*is told by*)
Domain: *HC14 Heritage Story*
Range: *HC17 Intangible Asset*

*HP24 includes visual documentation* (*visual documentation is included in*)
Domain: *HC3 Heritage Digital Twin*
Range: *HC6 Imagery*

*HP25 includes 3D documentation* (*3D documentation is included in*)
Domain: *HC3 Heritage Digital Twin*
Range: *HC7 3D Model*

*HP26 is digitization of* (*was digitised to create*)
Domain: *HC19 Heritage Digital Document*
Range: *HC18 Heritage Analogical Document*

*HP27 is digital part of* (*has digital part*)
Domain: *HC3 Heritage Digital Twin*
Range: *HC3 Heritage Digital Twin*

*HP28 is hosted by* (*hosts*)
Domain: *HC3 Heritage Digital Twin*
Range: *HP20 Hosting Service*

*HP29 is provided by* (*provides*)
Domain: *HP20 Hosting Service*
Range: *HP21 Hosting Service Provider*

*HP30 is titled* (*titles*)
Domain: *HC1 Heritage Entity*
Range: *HC12 Heritage Title*

*HP31 happened at location* (*was location for*)
Domain: *HC8 Heritage Event*
Range: *HC13 Heritage Location*

*HP32 was affected by* (*affected*)
Domain: *HC2 Heritage Asset*
Range: *crm:E5 Event*

*2.6 The ontology graph*

The following diagram (Figure 1) shows the structure of the HDT ontology with its most important classes and properties.

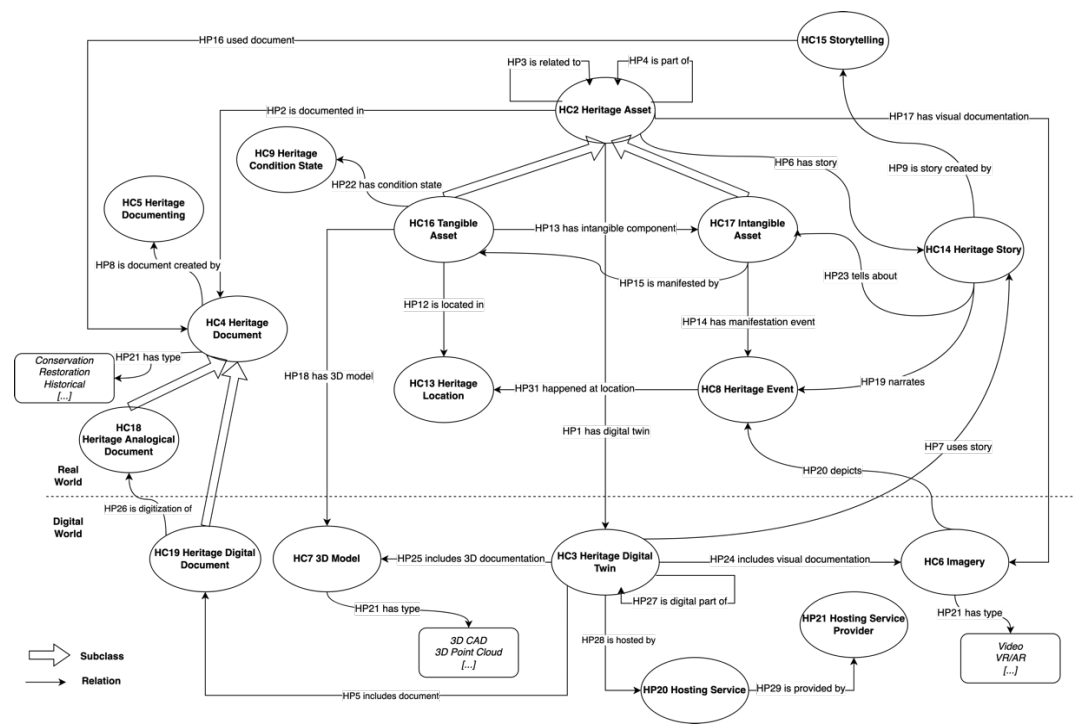

**Figure 1.** The semantic graph of the HDT ontology.

The structure of HC3 Heritage Digital Twin, a pivot concept tor the HDT ontology, and its related classes are outlined in Figure 2.

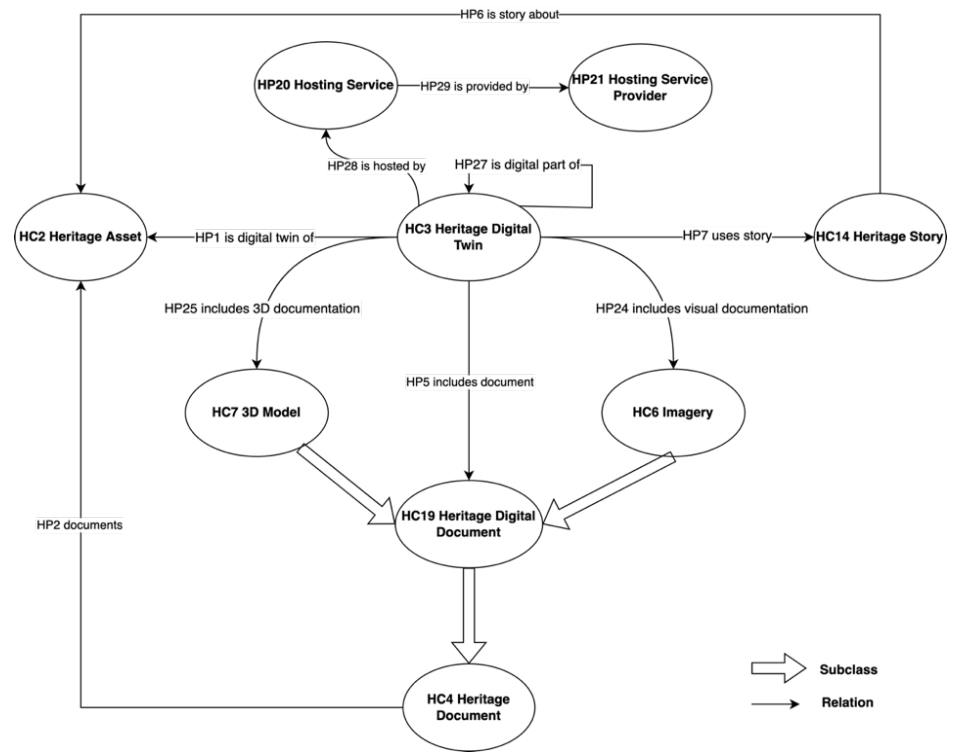

**Figure 2.** The semantic graph of *HC19 Heritage Digital Twin*.

## 3. Discussion

### 3.1 Suitability of the HDT ontology

The distinction used below for different aspects of cultural heritage by no means implies a strict classification of assets as belonging to any of the categories listed below. It rather analyses different facets of heritage, which may be present at the same time in the asset.

- Monuments appear to be well described by the HDT ontology.
- Sites are more difficult to deal with because of the vague definition of a "site". It may be an archaeological site, where remnants of the past were discovered and, in many cases, there are still standing structures. For archaeological sites, the corresponding digital twin must incorporate reference to all the discoveries made there, as they make the place a "site". It must also be capable of dealing with standing structures and their individual digital twins. Notable examples are Pompeii, where both assets, i.e. finds and standing structures, are present, and Stonehenge, where only the standing structures remain. Complex sites will be better represented by a global digital twin being the collection of the digital twins of the relevant parts of the site. In other cases, no structure survives, e.g. in very ancient sites, and only finds may contribute to the heritage digital twin. Thus, the site heritage digital twin consists of all relevant information as resulting from site investigations, and of the digital twins of surviving structures. In this case VR and AR models such as virtual reconstructions may store in a concise way plenty of information, besides being a help for visitors and an enhancement of the tourist attraction of the site. In this case, such models must comply with the requirements the London Charter [17], establishing principles for the computer visualisation of cultural heritage. Other examples of sites, potentially even more complicated, are historic centres. They include individual monuments but the characteristic of being a site stays with the complex, with its own digital twin, and often incorporates typical activities. Natural and cultural landscapes mix heritage and natural features which must be separately documented. An example from the UNESCO WH list [10], the landscape of Val d'Orcia in Tuscany, Italy, combines

- anthropogenic features with a particular natural and cultivated environment. Finally, other heritage sites are such only because of their history, as in the already mentioned case of battlefields that have neither remains nor finds.
- For movable artefacts, 3D models have a different significance depending on the nature of the object and from the way the multidimensional visual data are produced. 3D point-cloud models may be used to represent statues and other human-made objects, but for intrinsically 2-dimensional objects such as paintings or photos 2D imaging is in general better suited. 3D in the usual meaning of the term is in this case significant only in some special applications, for example to appreciate the paint technique in Pollock's *Alchemy* [18], or the deformation of the surface of Leonardo da Vinci's *The Adoration of the Magi* [19]. Other multidimensional imagery comes, among others, from scientific analyses such as tomography, layered non-invasive analyses, and so on. The resulting images supports interpretation and conservation activities. In some cases, AR / VR can help as well, when for example reconstructing the shape of an ancient musical instrument, accompanied with the sound it is supposed to have produced.
- Intangible heritage (and the intangible component of every heritage asset) is the most demanding component in terms of the HDT ontology development. As outlined above, it mainly relies on stories, and on the documentation of practices and individual or collective beliefs and traditions. To set up all the required classes and properties, we will start from what is already available here, and improve the fit by analysing and assessing the results of previous applications to intangible heritage.
- Born-digital heritage will have its own specifications, also part of the HDT ontology. The heritage digital twin will coincide in some parts with the asset, adding additional information to it.

As recommended in the 2021 Commission's Recommendation [1], special attention must be paid to so-called "High Value Datasets" (HVDS), which include the results of scientific analyses on artefacts for conservation or to improve the knowledge about the artefact under study, for example to date it. The semantic organization of such datasets has been studied in [20], [21], [22] and [23], and is compliant with the HDT ontology, into which it may be included.

In conclusion, the HDT ontology appears as a suitable foundation to build a knowledge base of cultural heritage, the main constituent of the heritage data space.

To incorporate legacy data in such knowledge base, mappings to the HDT ontology need to be produced as necessary. A possibility to be explored concerns the automatic production of parts of the HDT by means of artificial intelligence. Such enrichment includes, for example, text automatic annotation using Natural Language Processing (NLP), to recognize named entities in an unstructured text. An example of an application of such Named Entity Recognition (NER) is described in [23], where NER is applied to archaeological reports and to the documentation of scientific analyses. Another example discusses automatic segmentation of a 3D model to recognize its parts [24].

The next section outlines how to build a data space on the HDT ontology and exemplifies some services operating on it.

*3,2 Using the Heritage Digital Twin: the data space for cultural heritage as a knowledge graph*

The HDT ontology enables the creation of a heritage Knowledge Base (KB), where information about heritage is organized in a knowledge graph. Such graph can be managed and searched using a NoSQL DBMS, for example Ontotext GraphDB [25], which has already been applied in the cultural heritage domain. The Research Space project at the British Museum [26] is based on OntoText.

For instance, the Museo del Prado has developed a CIDOC CRM-based semantic graph for its collection [27]; the Rijksmuseum collections have bene organized as Linked Open Data [28]; and a semantic graph has been used for the creation of virtual exhibitions [29]. The main operation enabled by such DBs on the graph is searching the knowledge base and surfing it by means of Linked Data, where the links are represented by the common content for a class, i.e. by the same value for instances in that class and by properties relating classes with each other. With this approach, the heritage data space is not just a plain repository where cultural heritage data are stored and managed by a traditional database management system. Instead, it consists of the abstractions – the Heritage Digital Twins – of actual heritage assets, which model the real assets by selecting relevant features at the desired (or available) level of information granularity. In this regard, a compromise is reached between the detail, consisting of the raw data, and the abstraction, represented by the synthesis at ontology class level. For example, the material of a bronze object may be used as such, i.e. "bronze", to relate the object to other bronze objects; it may take into account the alloy composition, i.e. the percentage of copper, tin and other metals; in some cases it may refer to further details e.g. to classify the object provenance; finally, in very rare cases, the raw data of the analyses are used for linking and comparisons. All the required data are however stored in the heritage data space.

In conclusion, our vision of the data space for cultural heritage consists in a federation of (raw) data repositories distributed in distinct digital locations, possibly located in distinct geographical locations; and in a centralised system, the Knowledge Base (KB), which aggregates and manages synthetic information organised according to the HDT ontology approach. The KB content incorporates the metadata of the aggregated data, but also includes the relationships among them. This distributes the raw data storage load among repositories, operating mainly on the data synthesis stored in the KB by accessing raw data only when required.

The KB may be centralised in the same geographical place or split in distributed parts, all interconnected to make up the same knowledge base. The balance between centralization and decentralisation depends on the expected access and use: since cultural heritage is a location-based commodity, it may be worthwhile to evaluate the expected processing at the edges of the system, i.e. a use restricted to local data in a prevalent way, to optimise the efficiency of the global system through a balanced distribution of the data and the setup of edge computing processes. In any case, the federal approach for repositories also satisfies an aspect much appreciated by heritage institutions, i.e. the feeling that their data are kept under their control.

*3.2 Using the Heritage Digital Twin: automatic processing in the data space*

The heritage knowledge base is intended to be used by humans searching for information, but it is also the base for machine-actioned activities, e.g. triggered by sensors, which may send to a specific digital twin a continuous flow of signals or react to events or conditions requiring an intervention.

Under this regard, a digital twin is not a static concept, updated from time to time; it is instead a living replica of the real asset and is able to react autonomously, if necessary, according to specific conditions. Thus, sensors of humidity, for example, may trigger the system to limit the number of accesses to a monument, and an inexplicable strong increase in temperature may trigger a fire alarm and alert the fire brigade. Sensor data may also be stored for time-series analysis. Such processes require the use of AI systems, monitoring the HDT status to trigger actions on the real asset, warn custodians to act, or just record the event for future use. Thus, advanced digital technologies provide data to the cultural heritage data space, send it messages, and receive messages triggering for actions. The extent and effect of such reaction depend on how the simulation model analyses and reacts to external inputs.

As an example of automatic reaction, we can consider the following one.

The HDT system is connected to a system evaluating the risk of flood in some regions, based on meteo forecasts and a hydrogeological digital model. When such risk occurs, an alert activates a service, an automatic analyser searching for heritage asset locations placed in the risk region and comparing the expected flood height with asset basement height, calculated from the location and from the asset's 3D models. For those at risk of flooding, an analysis of the nature of the involved assets can indicate the most appropriate emergency measures, such as moving movable assets (e.g. paintings) at a higher position, putting barriers to water, and so on. As shown in Figure 3m instructions can then be sent to human officers in charge, and automatic systems may be activated when available, e.g. to raise barriers in the flood example.

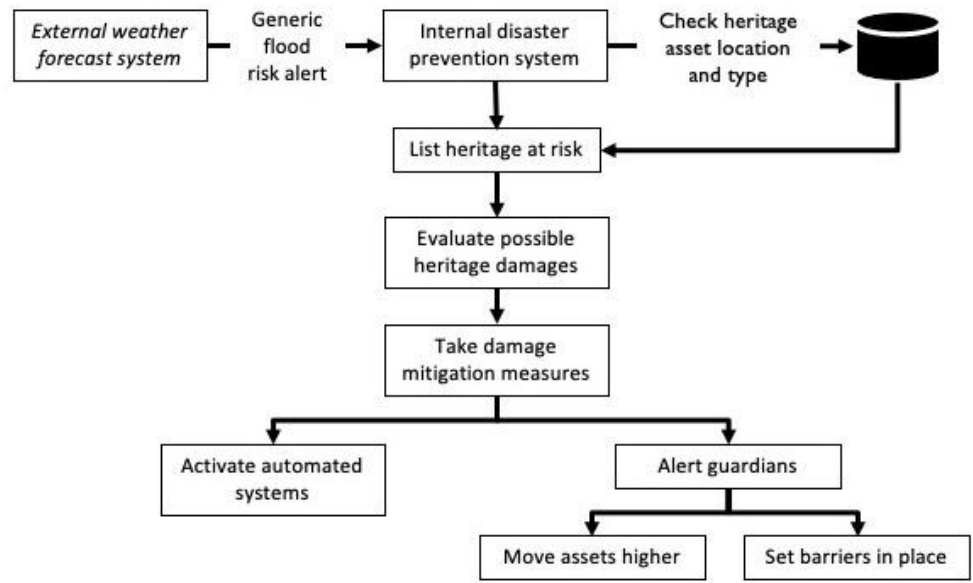

**Figure 3**. Schema of direct action activated by an automatic alert system based on HDT data.

For a real case, let us consider Cimabue's *Crucifixion* located at Santa Croce in Florence, Italy. The painting was badly damaged by the 1966 disastrous flooding of the local river Arno, and after a 50-years long restoration it has been placed back in its collocation, now with a lever system manually activated to raise the painting several meters up. Thus, when a top-level flood is feared basing on weather forecasts and the existing hydrogeological model of the Arno basin, a red alert is sent to the Civil Protection and forwarded – among others – to the Santa Croce custodians who run to manually activate the painting raising system. This procedure might be easily and effectively replaced with the automatic activation of such system triggered by the model prediction, automatic alarm phone calls to teams selected according to the evaluated risks and so on.

Another important opportunity provided by the heritage data space organization proposed here concerns the creation of Virtual Research Environments (VRE). A VRE is the digital simulation of a laboratory, where experiments are carried out on the digital representation of real items, i.e. on the digital twins, using software that simulates on the digital twins the effects of actions and events on their real counterpart. Thus, a VRE is the digital evolution of laboratories where real events are simulated in small-scale analogic models, such as a wind gallery or a ship model basin. Another feature of VREs is the collaboration they enable among different researchers who may access such shared digital spaces.

Examples concern the simulation of destruction caused by earthquake on monuments, or the impact of sea level raising due to global warming on monuments located at low height on the sea level, for example those in Venice.

*3,3 An example: Vasari's "Last Supper"*

In this section we develop a simple example to show how the HDT ontology may be used to create the digital twin of a heritage asset and to document some of its features, including a story regarding its location in different times and how this can be recorded availing of digital documentation.

Unfortunately, semantic information tends to be verbose, as it is conceived to be machine-processed and not read by humans. Therefore, we will not write the full semantic description as it would appear when codified in RDF or equivalent formats. Also the Turtle format [30] includes information going beyond the purpose of this example, so we will use a "simplified Turtle" enabling to write human-understandable statements omitting the information required for machine-processing it.

The simple rules of such description are the following ones:

- Classes and properties are written in italic.
- Instances of classes are written in bold and are followed by the class name they belong to. They are designated with short sentences to let the reader understand what they are, e.g. **The Pisa Leaning Tower** (*HC2 Heritage Asset*) means the instance of *HC2* which is known as the (famous) leaning tower.
- Properties relating classes are shown as relationships among instances. The subject, i.e. the instance of the property domain, is not repeated as long as it is the same. A change of subject is indicated by further indenting the statement. For example

    **The Pisa Leaning Tower** (*HC2 Heritage Asset*)
    *HP12 is located in* **Pisa** (*HC13 Heritage Location*)
    *HP2 is documented in* **Pisa Tourist Guide** (*HC4 Heritage Document*)
        *HP5 includes document* **Map of Pisa** (*HC4 Heritage Document*)

    means that the **Pisa Leaning Tower** is a heritage asset located in **Pisa.** It is described in the **Pisa Tourist Guide**. The latter includes a **Map of Pisa**.

- Any reference to the actual world, e.g. inventory numbers, addresses, DOI, etc. is used for the sake of the example and may, or may, not be a real one.
- Use of classes and properties directly taken from CIDOC CRM are indicated with *crm:* the namespace to which they belong. Note that *crm:D* and *crm:L* are used respectively for classes and properties defined in the CRMdig CRM extension [13].

The example concerns a 16th century painting by Giorgio Vasari titled "The Last Supper" and exhibited in the Santa Croce Museum in Florence. The painting was badly damaged by the 1966 flooding of the Arno River and after a long restoration it was placed back in this location. At the time of placing it back in the museum, a study was carried out to determine its original location in a nearby convent, the Murate Convent, from which it was removed in the 19th century following the Napoleonic convent suppression, confirmed by all the governments that ruled Tuscany afterwards and eventually turned the convent into a jail. The study was based on documents kept in the Florence State Archive, including architectural drawings and the 16th century handwritten diary of a nun reporting relevant details about the convent spaces. Such study enabled to reconstruct the whole original convent structure and the position of the refectory where the painting was originally placed, eventually leading to a reconstruction of the room as it was in the 16th century including the furniture and the painting in the original location. The details of this reconstruction process are reported in [31]. The reconstruction also produced a video with the story, currently visible on YouTube at the link included in the text below.

**S. Croce Church and Convent** (*HC16 Tangible Asset*)
  *HP12 is located in* **Piazza S. Croce, Florence, Italy** (*HC13 Heritage Location*)
  *HP4 is part of* **Florence Historic Centre** (*HC1 Heritage Entity*)
    *crm:P2 has type* "UNESCO WHS" (*crm:E55 Type*)
**Refectory in S. Croce Church and Convent** (*HC16 Tangible Asset*)
  *HP12 is located in* **Cloister of S. Croce Church, Piazza Santa Croce, Florence, Italy** (*HC13 Heritage Location*)
  *HP4 is part of* **S. Croce Church and Convent** (*HC16 Tangible Asset*)
**Murate Convent** (*HC16 Tangible Asset*)
  *HP12 is located in* **Via Ghibellina 2, Florence, Italy** (*HC13 Heritage Location*)
  *HP4 is part of* **Florence Historic Centre** (*HC1 Heritage Entity*)
    *crm:P2 has type* "UNESCO WHS" (*E55 Type*)

    **Painting Ultima Cena** (*HC16 Tangible Asset*)
        *HP30 is titled* "Ultima Cena" (*HC12 Heritage Title*)
        *HP30 is titled* "Last Supper" (*HC12 Heritage Title*)
        *HP10 is identified by* "1234-5678" (*HC 11 Heritage Identifier*)
        *HP12 is located in* **Refectory in S. Croce Church and Convent, Florence, Italy** (*HC13 Heritage Location*)
        *HP32 was affected by* **Florence Flooding of the Arno River in 1966** (*crm:E5 Event*)
        *crm:P31 was modified by* **Restoration of Vasari's Last Supper** (*crm:E11 Modification*)
            *crm:P14 carried out by* **Opificio delle Pietre Dure** (*crm:E39 Actor*)
        *HP2 is documented* in **Vasari's Last Supper Restoration** (*HC4 Heritage Document*)
        *HP22 has condition state* **Restored** (*HC9 Heritage Condition State*)
        *crm:P27* moved from **location of Murate Convent in Florence, Italy** (*HC13 Heritage Location*)
        *crm:P28* moved to **location of S.Croce Church and Convent in Florence, Italy** (*HC13 Heritage Location*)
        *crm:P28* moved to **refectory location within the S. Croce Church and Convent** (*HC13 Heritage Location*)
    **Digital Painting Ultima Cena** (*HC3 Heritage Digital Twin*)
        *HP1 is digital twin of* **Painting Ultima Cena** (*HC16 Tangible Asset*)
        *HP7 uses story* **The Original Location of Vasari's Ultima Cena** (*HC14 Heritage Story*)
        *HP25 includes 3D documentation* **3D Model of Murate Convent** (*HC7 3D Model*
    **Digital Murate Convent** (*HC3 Heritage Digital Twin*)
        *HP1 is digital twin of* **Murate Convent** (*HC16 Tangible Asset*)
        *HP5 includes document* **Scan of 1826 Elevation of the Murate convent** (*HC19 Heritage Digital Document*)
            *HP26 is digitization of* **1826 Elevation of the Murate convent** (*HC18 Heritage Analogical Document*)
                *HP10 is identified by* "Florence State Archive (ASF), SFF, FL, 2109, II, ins.132bis, N35" (*HC11 Heritage Identifier*)
                *HP21 has type* "Architectural Elevation" (*crm:E55 Type*)
        *HP5 includes document* "**Scan of 1837 Plan of the Murate convent** (*HC19 Heritage Digital Document*)
            *HP26 is digitization of* **1837 Plan of the Murate convent** (*HC18 Heritage Analogical Document*)
                *HP10 is identified by* "Florence State Archive (ASF), FL, 12399" (*HC11 Heritage Identifier*)
                *HP21 has type* "Architectural Plan" (*crm:E55 Type*)
        *HP5 includes document* **Digital Diary of Suor Giustina** (*HC19 Heritage Digital Document*)
            *HP26 is digitization of* **Diary of Suor Giustina** (*HC18 Heritage Analogical Document*)
                *HP10 is identified by* "Florence State Archive folder 32399" (*HC11 Heritage Identifier*)
                *HP21 has type* "Handwritten diary" (*crm:E55 Type*)
        *HP25 includes 3D documentation* **3D Model of Murate Convent** (*HC7 3D Model*)
        *HP7 uses story* **Ultima Cena Original Location** (*HC14 Heritage Story*)
            *HP9 is story created by* **Ultima Cena Storytelling** (*HC15 Storytelling*)
                *HP16 used document* **Youtube video https://www.youtube.com/watch?v=P1Uv4Zf5xKk** (*HC19 Heritage Digital Document*)
                *HP16 used document* **3D Model of Murate Convent** (*HC7 3D Model*)
        *HP6 is story about* **Painting Ultima Cena** (*HC16 Tangible Asset*)
    **3D Model of Murate Convent** (*HC7 3D Model*)
        *crm:P2 has type* **Reconstruction 3D Model** (*E55 Type*)
        *crm:L11 was output of* **Creation of Murate 3D model** (*crm:D7 Digitization Machine Event*)

 *crm:L23 used software* **3ds Max** (*crm:D14 Software*)
 *crm:L10 had input* **Scan of 1826 Elevation of the Murate convent** (*HC19 Heritage Digital Document*)
 *crm:L10 had input* **Scan of 1837 Plan of the Murate convent** (*HC19 Heritage Digital Document*)
 *crm:L10 had input* **Digital Diary of Suor Giustina** (*HC19 Heritage Digital Document*)

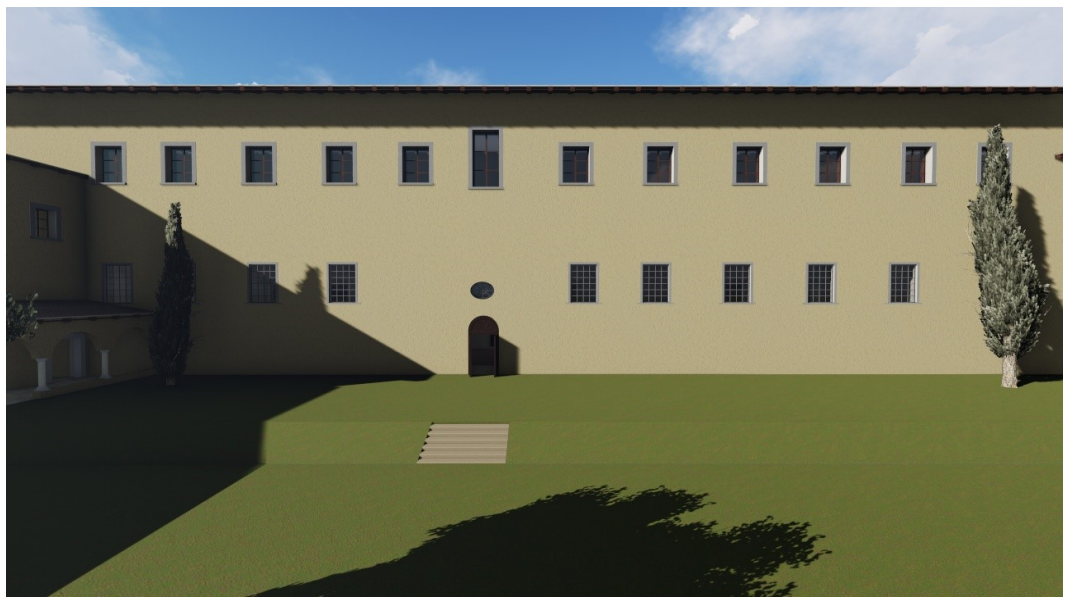

Figure 4. The 3D reconstruction of the Murate Convent inner garden. N. Amico & F.Niccolucci 2016, CC-BY.

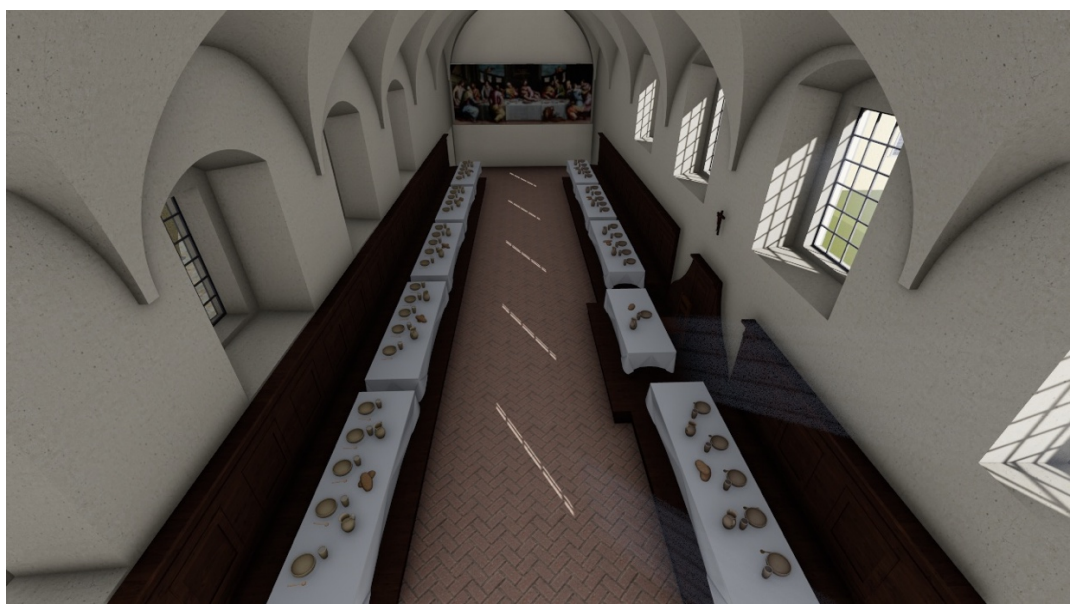

Figure 5. The 3D reconstruction of the Murate Convent Refectory as it appeared in the 16th century, with the *Last Supper* painting at the end of the room. N. Amico & F.Niccolucci 2016, CC-BY.

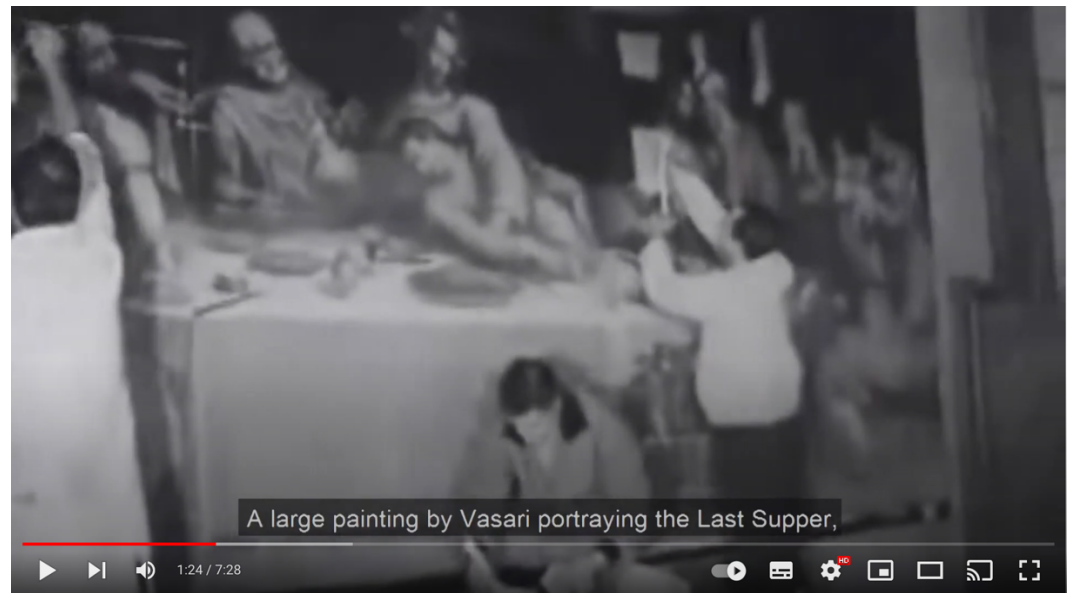

**Figure 6**. A screenshot from the YouTube video with 1966 black and white TV footage documenting emergency interventions on the painting. PRISMA 2016, CC-BY.

*3.4 A case study: the Pafos Gate in Nicosia, Cyprus and its heritage digital twin*

The Pafos gate (or "The Upper Gate", being at a slightly higher elevation that the others, at ca. 150 m. above the sea level) is one of the three accesses to the walled old city of Nicosia. It connects the access routes to the city from the west, through the main commercial area of the city, with the routes towards east, accessible via the Famagusta gate. Its history can be traced back to the Venetian administration of Francesco Barbaro, Proveditore (military administrator of the kingdom), when the Italian military engineer Giulio Savorgnan was assigned the role of building the defense system of the capital, in light of the imminent ottoman threats. He designed and supervised the construction of the defense system for eight months in 1567, which continued afterwards under the guidance of his assistant Leonardo Roncone until 1568.

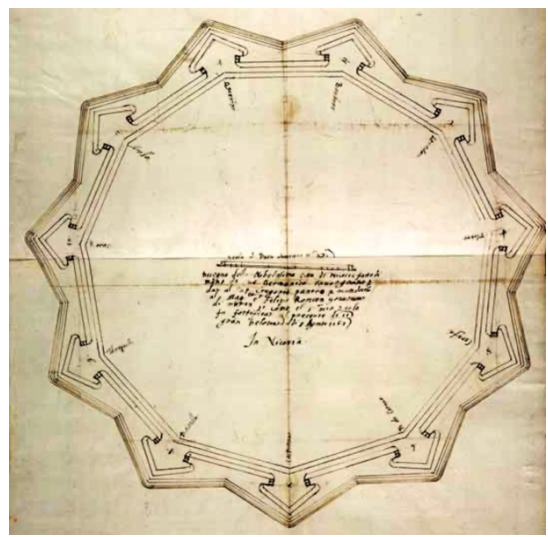

**Figure 7**. Planning the Nicosia fortress by Giulio Savorgnan: sketch of the Nicosia enceinte [32].

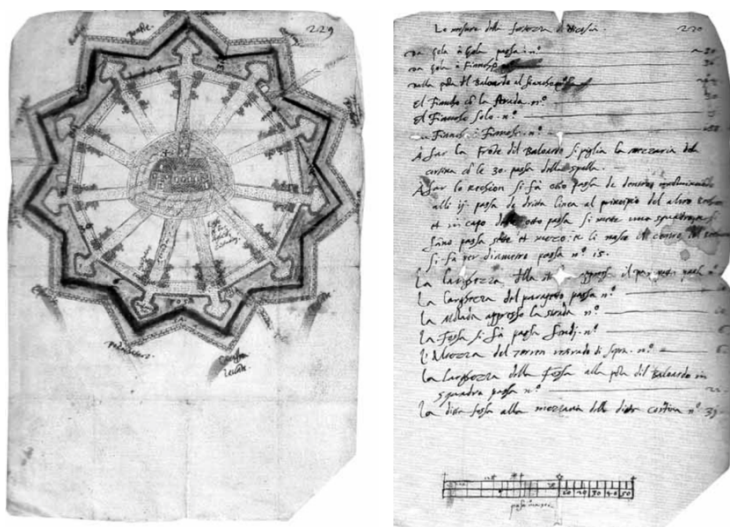

**Figure 8.** Planning the Nicosia fortress by Giulio Savorgnan: measurements of the fortress and detailed description and measurements of each enceinte component [32].

The gate complex has two broad staircases leading from its internal access to the walls. A covered sewage system runs along the edges of the access tunnel. The gate complex is illustrated in Figure 9.

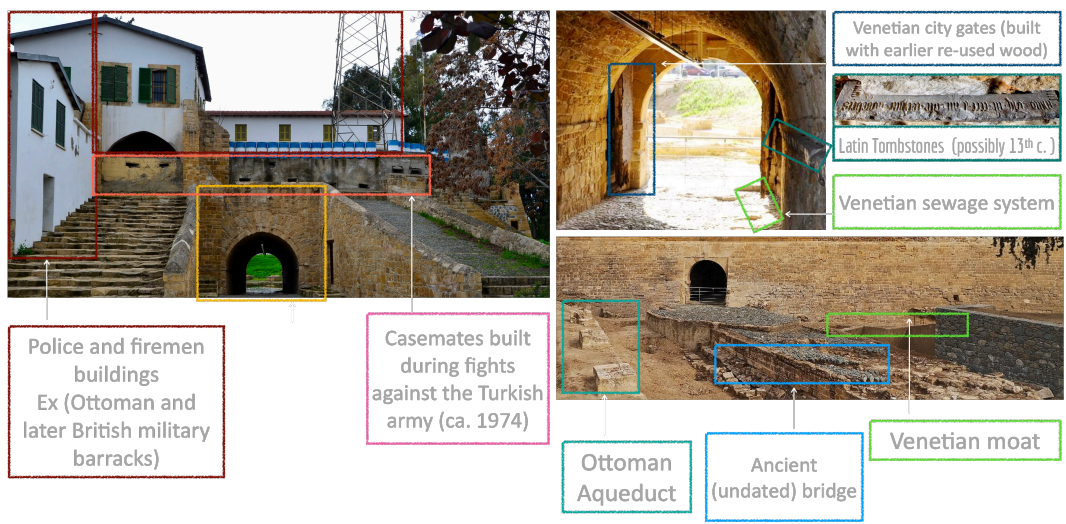

**Figure 5**. The Pafos gate with its main components.

The riverbed of the Pedeios river, in earlier periods crossing the city, was diverted so as its waters would fill the stellar-shaped moat, ca. 80 m. wide, surrounding the city. Earlier medieval fortifications, along with churches and buildings from the same period located across the perimeter of the new defense wall (such as the San Domenico monastery), were demolished, stones being re-used in the construction of the walls. These consist of a hendecagonal wall primarily made of earth filling and stone cover, ca. 7 km. in diameter and eleven bastions of similar size and shape located at ca. 260 m. intervals one from the other. The walls were built with a slope at an angle as such as to reduce as much as possible the forces of the bullets. The rampart consists of earth, with a facing of stone halfway up its height forming a retaining wall for the lower portion. Above this retaining wall the scarped face of the earthwork was intended to form a grassy slope [33]. Whereas early modern photographs show that the gate tunnel had wooden doors on both its ends [34], nowadays only one is in place (at the external end of the tunnel) while the fate of the other is unknown. The surviving door is made of vertical beams of wood (predating the Venetian period) reinforced by horizontal metal long plates nailed into the wood with iron spikes (Figure 5).

During the Ottoman rule on the island (1571 – 1878), the walls of Nicosia were reinforced - a barely discernible inscription on the gate's internal arch, possibly from the reign of Sultan Mahmut the 2$^{nd}$ (ca. 1821) was added on the wall just above the entrance to the gate's access tunnel; military barracks were built on the walls above and adjacent to the gate. Recent excavations just outside the gate revealed, among others, remains of the Venetian mote, a bridge and parts of the Ottoman period Arab Ahmet Pasha aqueduct. The quarter inside the walls and in from the gate were known as Tabakhane (i.e., Tannery) Quarter. Further changes occurred during the British Protectorate of the island (1878 – 1914), when the ground inside the gate was raised by circa half a meter to avoid flooding, the gate was closed and a breach in the wall next to the gate was opened so as to allow increased traffic into the city. A last addition on top of the gate is a series of concrete casemates built in haste during the fights against the invading Turkish troops in 1974. In recent years, as part of an initiative to re-open the gate as part of a cultural tourism circuit of the city, excavations were held in front of the gate and an access bridge was constructed over this area and connect the sidewalk with the gate, above the excavated features.

The above information is encoded below, using the HDT ontology. The first list includes the value of the instances used in the encoding, which is presented below using again the simplified Turtle. As the descriptions of the instances are relevant here – it is a real case – they are provided separately to avoid messing up the Turtle encoding.

**Pafos Gate** (*HC16 Tangible Asset*)
    *HP1 has digital twin* **Pafos Gate Digital Twin** (*HC3 Heritage Digital Twin*)
    *HP2 is documented in* **Pafos Gate Documentation Folder** (*HC4 Heritage Document*)
    *HP2 is documented in* **Giulio Savorgnan's Letters** (*HC18 Heritage Analogical Document*)
    *HP2 is documented in* **Giulio Savorgnan Drawings** (*HC18 Heritage Analogical Document)*
    *HP2 is documented in* **Fra Stefano Lusignano's Chronography** (*HC18 Heritage Analogical Document*)
    *HP3 is related to* **Nicosia Venetian Fortification System** (*HC16 Tangible Asset*)
    *HP4 is formed by part* **Pafos Gate Doors** (*HC16 Tangible Asset*)
    *HP4 is part of* **Nicosia Venetian Fortification System** (*HC16 Tangible Asset*)
    *HP4 is part of* **Police/Firemen Headquarters** (*HC16 Tangible Asset*)
    *HP6 has story* **Pafos Gate Story** (*HC14 Heritage Story*)
    *HP10 is identified by* "DoAIN2022" (*HC11 Heritage Identifier*)
    *HP30 is titled* "The Pafos Gate" (*HC12 Heritage Title*)
    *HP30 is titled* "High Gate" (*HC12 Heritage Title*)
    *HP30 is titled* "Porta San Domenico" (*HC12 Heritage Title*)
    *HP12 is located in* **Nicosia, Cyprus** (*HC13 Heritage Location*)
    *HP13 has intangible component* **Tanners and Dyers Commerce** (*HC17 Intangible Asset*)
        *HP14 has manifestation event* **Activity of Tanners and Dyers Workshops** (*HC8 Heritage Event*)
            *HP31 happened at location* **Nicosia, Cyprus** (*HC13 Heritage Location*)
    *HP17 has visual documentation* **Pafos Gate Tunnel Image** (*HC6 Imagery*)
    *HP18 has 3D model* **Pafos Gate Laser Scanning** (*HC7 3D Model*)

**Pafos Gate Digital Twin** (*HC3 Heritage Digital Twin*)
    *HP5 includes document* **Pafos Gate Documentation** (*HP19 Heritage Digital Document*)
        *HP8 is document created by* **Pafos Gate Documenting Activity** (*HC5 Heritage Documenting*)

  *HP5 includes document* **Scan of Giulio Savorgnan's Letters** (*HP19 Heritage Digital Document*)

    *HP26 is digitization of* **Giulio Savorgnan's Letters** (*HC18 Heritage Analogical Document*)

  *HP7 uses story* **Pafos Gate Story** (*HC14 Heritage Story*)

  *HP24 includes visual documentation* **Pafos Gate Tunnel Image** (*HC6 Imagery*)

  *HP25 includes 3D documentation* **Pafos Gate Laser Scanning** (*HC7 3D Model*)

    *HP21 has type* **3D Point Cloud** (*crm:E55 Type*)

  *HP28 is hosted by* **STARC Repository** (*HC20 Hosting Service*)

    *HP29 is provided by* **The Cyprus Institute** (*HC21 Hosting Service Provider*)

**Pafos Gate Story** (*HC14 Heritage Story*)

  *HP9 is story created by* **Correspondence between Giulio Savorgnan and Venetian Officials** (*HC15 Storytelling*)

    *HP16 used document* **Giulio Savorgnan's Letters** (*HC18 Heritage Analogical Document*)

  *HP9 is story created by* **Fra Stefano Lusignano's Narration** (*HC15 Storytelling*)

    *HP16 used document* **Fra Stefano Lusignano's Chronography** (*HC18 Heritage Analogical Document*)

  *HP9 is story created by* **Giulio Savorgnan's fortifications building plan** (*HC15 Storytelling*)

    *HP16 used document* **Giulio Savorgnan Drawings** (*HP18 Heritage Analogical Document*)

  *HP9 is story created by* **Reconstruction of Pafos Gate Uses** (*HC15 Storytelling*)

    *HP16 used document* **Pafos Gate Documentation** (*HP19 Heritage Digital Document*)

    *HP16 used document* **Pafos Gate Architectonic Documentation** (*HP19 Heritage Digital Document*)

      *HP8 is document created by* **Pafos Gate 3D Modelling Activity** (*HC5 Heritage Documenting*)

**Instances, Classes and Descriptions**

(boldface denotes Class instances; text within "" is a character string)

Instance: **Pafos Gate Digital Twin**
Instance of Class: *HC3 Heritage Digital Twin*
Instance Description: http://public.cyi.ac.cy/starcRepo/explore/objects Pafos Gate collection

Instance: **Pafos Gate Documentation Folder**
Instance of Class: *HC4 Heritage Document*
Instance Description: Folder containing all material on the study of the gate in order to develop a 3D virtual environment, as part of the rehabilitation plan of the Nicosia municipality to transform the area into a visitable archaeological park.

Instance: **Pafos Gate Documenting Activity**
Instance of Class: *HC5 Heritage Documenting*
Instance Description: The activity of collecting or digitizing documents concerning the Pafos Gate

Instance: **Pafos Gate 3D Modelling Activity**
Instance of Class: *HC5 Heritage Documenting*
Instance Description: Activity of creating 3D models of the various architectonic components of the Gate

Instance: **Pafos Gate Tunnel Image**
Instance of Class: *HC6 Imagery*

Instance Description: *http://public.cyi.ac.cy/starcRepo/details/show /2d131880fdb75dbcb3c35cf8df74b1ca*

Instance: **Pafos Gate Laser Scanning**
Instance of Class: *HC7 3D Model*
Instance Description: *http://public.cyi.ac.cy/starcRepo/details/show/5a628b796fc6100f5a3a2e2dce1dc175*

Instance: **Activity of Tanners and Dyers Workshops**
Instance of Class: *HC8 Heritage Event*
Instance Description: Specific events manifesting the activity of the Tanners and Dyers Workshop near the Pafos Gate

Instance: **REVEEL 3D**
Instance of Class: *HC10 Heritage Event Model*
Instance Description: https://avl.ncsa.illinois.edu/realworld-software/riveel-3d

Instance: **DoAIN2022**
Instance of Class: *HC11 Heritage Identifier*
Instance Description: inventory number of the gate in the Department of Antiquities catalogue of monuments.

Instance: **"The Pafos Gate"**
Instance of Class: *HC12 Heritage Title*
Instance Description: The Pafos Gate (ex. High Gate, earlier Porta San Domenico)

Instance: **Nicosia Cyprus**
Instance of Class: *HC13 Heritage Location*
Instance Description: Nicosia, Cyprus 35.1737° N, 33.3568° E

Instance: **Pafos Gate Road**
Instance of Class: *HC13 Heritage Location*
Instance Description: Road which leads to the Pafos Gate from the centre of the city, part of the commercial route that crossed the city connecting the Pafos gate with the Famagusta gate.

Instance: **Pafos Gate Story**
Instance of Class: *HC14 Heritage Story*
Instance Description: Story of the Pafos Gate throughout the centuries

Instance: **Correspondence between Giulio Savorgnan and Venetian Officials**
Instance of Class: *HC15 Storytelling*
Instance Description: Story of the correspondence between Giulio Savorgnan and Venetian officials

Instance: **Giulio Savorgnan's fortifications building plan**
Instance of Class: *HC15 Storytelling*
Instance Description: Giulio Savorgnan's plans to build the fortifications

Instance: **Fra Stefano Lusignano's Narration**
Instance of Class: *HC15 Storytelling*
Instance Description: Storytelling of the Chronography of the Cyprus island by Fra Stefano Lusignano

Instance: **Reconstruction of Pafos Gate Uses**
Instance of Class: *HC15 Storytelling*
Instance Description: Reconstruction of the history of the various uses of the Pafos gate throughout the centuries

Instance: **Pafos Gate**
Instance of Class: *HC16 Tangible Asset*
Instance Description: The Pafos Gate

Instance: **Nicosia Venetian Fortification System**
Instance of Class: *HC16 Tangible Asset*

Instance Description: Nicosia Venetian fortification system; commercial routes from the west to the east

Instance: **Police/Firemen Headquarters**
Instance of Class: *HC16 Tangible Asset*
Instance Description: Police / Firemen Headquarters; part of a new cultural tourism path in Nicosia

Instance: **Pafos Gate Doors**
Instance of Class: *HC16 Tangible Asset*
Instance Description: The wooden doors enforced by iron beams.

Instance: **Tanners and Dyers Commerce**
Instance of Class: *HC17 Intangible Asset*
Instance Description: Tanners and Dyers workshops were located along the road which leads to the Pafos Gate from the centre of the city.

Instance: **Giulio Savorgnan Drawings**
Instance of Class: *HP18 Heritage Analogical Document*
Instance Description: Drawings of the fortifications made by Giulio Savorgnan.

Instance: **Giulio Savorgnan's Letters**
Instance of Class: *HC18 Heritage Analogical Document*
Instance Description: Collection of letters written by Giulio Savorgnan to Venetian officials, describing the design and building plans of the Venetian fortifications (Pafos gate part of them)

Instance: **Fra Stefano Lusignano's Chronography**
Instance of Class: *HC18 Heritage Analogical Document*
Instance Description: The *Chorograffia et breve historia universale dell'isola de Cipro* of Fra Stefano Lusignano, a Dominican friar from the royal family. Printed first in Bologna, 1573.

Instance: **Scan of Giulio Savorgnan's Letters**
Instance of Class: *HP19 Heritage Digital Document*
Instance Description: Digital collection of letters written by Giulio Savorgnan to Venetian officials, describing the planning and building of the Venetian fortifications (Pafos gate being part of them) and including measurements, plan drawings and textual descriptions.

Instance: **Pafos Gate Documentation**
Instance of Class: *HP19 Heritage Digital Document*
Instance Description: D documentation of the gate, RTI on specific details (the inscription, the lapidary stones), dendrochronology analysis of the wooden beams of the doors.

Instance: **Pafos Gate Architectonic Documentation**
Instance of Class: *HP19 Heritage Digital Document*
Instance Description: D models of the various architectonic components of the Gate

Instance: **STARC Repository**
Instance of Class: *HC20 Hosting Service*
Instance Description: STARC Repo - http://public.cyi.ac.cy/starcRepo/

Instance: **The Cyprus Institute**
Instance of Class: *HC21 Hosting Service Provider*
Instance Description: The Cyprus Institute (CYI)

## 5. Conclusions

The ontology introduced in the present paper needs further refinement, which may only come from testing it in practice and getting feedback by the heritage community and stakeholders. It is also particularly important to refine the documentation of relevant features concerning for instance the quality and detail of 3D models, the procedure and settings used in scientific analyses, and so on. On this regard, ongoing work within the ARIADNEplus project concerning application profiles and the 4CH project especially as regards 3D modelling and conservation activities provide components that already usable within the HDT ontology.

The HDT ontology is primarily aimed at organizing and managing digital information about heritage, in order to produce what the EU Commission calls "high quality records" in the *Recommendation on a common European data space for cultural heritage* [1], for which it envisages an increase of about 60% by 2025 – a very ambitious target indeed. It is clear that high quality records require rich, high quality metadata, such as those implemented by the HDT ontology proposed here. It relies on and is fully compliant with the CIDOC CRM international standard and its compatible models, as the one adopted by Europeana in the EDM ontology [14] and those adopted by a large and increasing number of galleries, museums, libraries and other cultural institutions. It is also compatible with the ARIADNE catalogue of about two million archaeological datasets and the forthcoming 4CH knowledge base. Actually, the EDM is equivalent to a subset of HDT so the improvement of the quality of existing records, also recommended by the Commission, is progressively feasible with disrupting the compatibility with the current wealth of records provided by Europeana.

With such rich metadata, the system will enable users to perform the activities typical of working with cultural heritage, including the access and retrieval of information and its use to know, understand, preserve, communicate and share the assets. Adopting this scheme and its semantic graph guarantees retrievable and accessible data. Using the HDT ontology enables interoperability and reuse, making heritage data FAIR. It supports the creation of a distributed system, a federation of storage and cloud facilities concerning heritage across Europe, as indicated by the EU Commission for the Data Space for Cultural Heritage in its Communication [1]. It enables linking with other economic sectors such as tourism and creative industries. It supports the development of advanced services, as sketched in section 3.2. It is ready to link or to incorporate related initiatives such as the forthcoming Competence Centre for Cultural Heritage to be designed by the 4CH project [16] focusing on a digital approach to conservation and preservation. It enables collaboration among institutions and professionals, as well as with citizens, via the establishment of VREs. In sum, it has all the characteristics required for a starting point when building a cultural heritage data space as envisaged by European strategies.

**Author contributions**. All authors have equally contributed to all the activities leading to the preparation of the present paper. All authors have read and agreed to the published version of the manuscript.

**Funding**. The research received no external funding.

**Conflict of interest**. Authors declaee there is no conflict of interest

*Data* **2022**, *7*, x FOR PEER REVIEW 28 of 56. ICOM. 2017. *Statute*. Available at https://icom.museum/wp-content/uploads/2018/07/2017_ICOM_Statutes_EN.pdf.
7. Singh, M.; Fuenmayor, E.; Hinchy, E.P.; Qiao, Y.; Murray, N.; Devine, D. Digital Twin: Origin to Future. *Appl. Syst. Innov.* 2021, *4*, 36. https://doi.org/10.3390/asi4020036
8. ISO. 2018. ISO 16739-1:2018. Available at: https://www.iso.org/standard/70303.html.
9. Bolton, A.; Lorraine, B.; Dabson, I.; Enzer, M.; Evans, M; Fenemore, T.; Harradence, F.; Keaney, E.; Kemp, A.; Luck, A.; Pawsey, N.; Saville, S.; Schooling, J.; Sharp, M.; Smith, T.; Tennison, J.; Whyte, J.; Wilson, A.; Makri, C. *The Gemini Principles: Guiding values for the national digital twin and information management framework.* 2017 https://doi.org/10.17863/CAM32260.
10. UNESCO. *World Heritage List*. Available at https://whc.unesco.org/en/list
11. UNESCO. *Intangible Cultural Heritage List*. Available at https://ich.unesco.org/en/lists.
12. Niccolucci, F. 2007. Virtual museums and archaeology: an international perspective. *Archeologia e Calcolatori*, 2007, *17*, pp. 15-30.
13. CIDOC-CRM. 2021. http://www.cidoc-crm.org/
14. EuropeanaPro. The Europeana Data Model. 2017. https://pro.europeana.eu/page/edm-documentation.
15. ARIADNE. Web site: https://ariadne-infrastructure.eu/.
16. 4CH. Web site: https://4ch-project.eu/.
17. The London Charter. Web site: https://The-London-Charter.org/
18. Callieri, M.; Dellepiane, M.; Pavoni, G.; Pingi, P.; Potenziani, M.; Scopigno, R. Alchemy in 3D: A digitization for a journey through matter. Proceedings of the International Congress on Digital Heritage, Granada, Spain, 28/08/2015-02/10/2015. 2015, pp. 223–231. DOI: 10.1109/DigitalHeritage.2015.7413875.
19. Palma, G.; Pingi, P.;Siotto, E.;Bellucci, R.;Guidi, G.; Scopigno, R. Deformation analysis of Leonardo da Vinci's "Adorazione dei Magi" through temporal unrelated 3D digitization. *Journal of Cultural Heritage* 2109, *38*, pp. 173-185.
20. Felicetti, A. ; Niccolucci, F.: A CIDOC CRM-based Model for the Documentation of Heritage Sciences. Proceedings of thr 3rd International Congress on Digital Heritage (DH2018). Institute of Electrical and Electronics Engineers, 2018. DOI: 10.1109/DigitalHeritage.2018.8810109
21. Castelli, L.; A. Felicetti, A. Modelli e strumenti semantici per la documentazione dell'Heritage Science, Atti della Conferenza GARR 2018, Cagliari, 3-5 Ottobre*. 2018.*, https://www.garr.it/it/news-e-eventi/pubblicazioni/atti-delle-conferenze/atti-della-conferenza-garr-2018
22. Castelli, L.; Felicetti, A.; Proietti, F. Heritage Science and Cultural Heritage: a CIDOC-CRM-enabled model for Integration and Interoperability", *International Journal on Digital Libraries, Special issue on FAIR Data and Cultural Heritage data-centric research,* 2019. DOI: https://doi.org/10.1007/s00799-019-00275-2
23. Bombini, A.; Castelli, L.; dell'Agnello, L.; Felicetti. A.; Giacomini, F.; Niccolucci, F.; Taccetti, F. CHNet cloud: an EOSC-based cloud for physical technologies applied to cultural heritages. Proceedings of the GARR Conference 2021, Online, 7-16 June 2021. 2021. DOI: 10.26314/GARR-Conf21-proceedings-09.
24. Gonizzi Barsanti, S.; Guidi, G.; De Luca, L. Segmentation of 3D models for cultural heritage structural analysis – some critical issues. *ISPRS Annals of the Photogrammetry, Remote Sensing and Spatial Information Sciences,* 2017 *Volume IV-2/W2*, Proceedings of the 2017 - 26th International CIPA Symposium 2017, pp. 115-122.
25. Ontotext. Web site: https://graphdb.ontotext.com/.
26. Research Space. Web site: https://researchspace.org/
27. Museo del Prado *The Museo del Prado's Ontological Model* Available at: https://www.museodelprado.es/en/grafo-de-conocimiento/modelo-ontologico.